\documentclass{aa}

\usepackage{xspace}
\usepackage{amssymb}
\usepackage{amsfonts}
\usepackage{epsfig}

\begin{document}
\thesaurus{}
\title{Kinematics of LMC stellar populations and
self-lensing optical depth}
\author{
P.~Salati\inst{1,2},
R.~Taillet\inst{1,2}
{\'E}.~Aubourg\inst{3},
N.~Palanque-Delabrouille\inst{3},
M.~Spiro\inst{3},
}
\institute{
LAPTH, chemin de Bellevue, BP 110, 74941 Annecy-le-Vieux Cedex, France.
\and Universit{\'e} de Savoie, B.P. 1104, 73011 Chamb{\'e}ry Cedex, France.
\and CEA, DSM, DAPNIA,
Centre d'{\'E}tudes de Saclay, 91191 Gif-sur-Yvette Cedex, France
}
\offprints{Eric.Aubourg@cea.fr}
\date{Received;accepted}
\authorrunning{Salati et al.}
\titlerunning{Kinematics of {\sc lmc} stellar populations and self--lensing}

\maketitle

\def\eros{{\sc eros}\xspace}
\def\macho{{\sc macho}\xspace}
\def\lmc{{\sc lmc}\xspace}
\def\smc{{\sc smc}\xspace}
\def\etal{{et al.}\xspace}
\def\tauLMC{\tau_{\mathrm{\mathsc{lmc}}}}
\def\DLMC{D_{\mathrm{\mathsc{lmc}}}}

\newcommand{\unit}[1]{\; \mbox{#1}}
\newcommand{\un}[1]{_{\scriptsize \mbox{#1}}}
\newcommand{\Msol}{{\rm M}_\odot}
\newcommand{\Lsol}{{\rm L}_\odot}
\newcommand{\SW}{\mbox{$\sigma_{W}$}}
\newcommand{\beq}{\begin{equation}}
\newcommand{\eeq}{\end{equation}}
\newcommand{\ie}{{\it i.e.}}
\newcommand{\hepex}[1]{{\tt hep-ex/#1}}
\newcommand{\hepph}[1]{{\tt hep-ph/#1}}
\newcommand{\astroph}[1]{{\tt astro-ph/#1}}

\begin{abstract}

Recent observations give some clues that the lenses discovered by the 
microlensing experiments in the direction of the Magellanic Clouds may 
be located in these satellite galaxies.  We re-examine the possibility 
that self-lensing alone may account for the optical depth measured 
towards the Large Magellanic Cloud (\lmc).  We present a 
self-consistent multi-component model of the \lmc consisting 
of distinct stellar populations, each associated to a vertical velocity 
dispersion ranging from 10 to 60 km/s.  The present work focuses on 
showing that such dispersions comply with current $20 - 30$ km/s limits set 
by observation on specific \lmc populations.  We also show that this 
model reproduces both the $1 - 2 \times 10^{-7}$ observed optical 
depth and the event duration distribution.

\end{abstract}

Several collaborations (Alcock et al. 1993, Au\-bourg et al. 1993) are
searching for galactic dark matter through the use of gravitational
microlensing (Paczy\'{n}ski 1986) towards the Magellanic
Clouds. Events have been observed, for which location and mass
cannot be determined independently. The current results do not yet
yield a coherent explanation: half of the halo of the Milky Way in
0.5~M$_{\odot}$ objects (\cite{Macho2yr}) would require a puzzling star
formation history, whereas traditional models of the \lmc do not
predict a self-lensing optical depth high enough to account for all
the observed events (Gould 1995). The only events with additional
information all seem to be located in the Clouds themselves
(Bennett et al. 1996, Palanque-Delabrouille et al. 1998, Afonso et al.
1999), which makes it worthwhile to re-examine the experimental
constraints on the Clouds kinematics and explore more thoroughly
models of the \lmc.
After reviewing the observational constraints on the \lmc kinematics 
(section 1), we show, in section 2, the existence of an age bias: the 
stars used to derive these constraints are on average both younger and 
slower than the majority of the \lmc objects.  We then use a Monte 
Carlo simulation to show that a maximum velocity dispersion of 60 km/s 
reproduces the kinematic observations (section 3) and the microlensing 
results (section 4).

\section{Present observational constraints}

The bulk of the mass of the \lmc resides in a nearly face-on disk,
with an inclination usually taken to equal the canonical value of
$i=33^{\circ}$ (\cite{westerlund}), although both lower ($27^{\circ}$)
and higher (up to $45^{\circ}$) values have also been derived from
morphological or kinematical studies of the \lmc. This disk is
observed to rotate with a circular velocity $V_{C} \sim 80$~km/s out
to at least $8^{\circ}$ from the \lmc center
(\cite{schommer}).
If all the stars belong to the same population, with a vertical ({\it 
i.e.} perpendicular to the disk) velocity dispersion $\sigma_{W}$, the 
microlensing optical depth of such a disk upon its own stars is given 
by $\tau \sim 2 \sigma_{W}^{2} \sec^{2}i / c^{2} $ (\cite{gou95}).  
Considering the measured velocity of \lmc carbon stars (Cowley \& 
Hartwick 1991), Gould (1995) assumed $\sigma_{W} = 20$ km/s as 
a typical velocity dispersion for \lmc stars.  He thus concluded that 
$\tau \sim 10^{-8}$, {\it i.e.} that self-lensing (first suggested by 
\cite{sahu94} and \cite{wu94}) contributes very little to the observed 
optical depth towards this line of sight.

Carbon stars however may not be the ultimate probe to infer the velocity
dispersion of \lmc populations: they actually comprise various
ill-defined classes of objects (Me\-nes\-sier 1999), and their prevalence 
is a complex function of age, metallicity and probably other factors 
(Gould 1999). 

Both observational and theoretical arguments favour the existence
of a wide range of velocity dispersions among the various \lmc stellar
populations.
To commence, Mea\-theringham \etal (1988) have determined the
radial velocities of a sample of planetary nebulae (PN) in the \lmc.
They measured a velocity dispersion of 19.1 km/s, much larger than the value
of 5.4 km/s found for the HI. This was interpreted as being
suggestive of orbital heating and diffusion operating in the \lmc in
the same way as it is observed in the solar neighbourhood.
Then, the observations of Hughes \etal (1991) show clear
evidence for an increase in the velocity dispersion of long period
variables (LPV) as a function of their age. For young LPVs, the velocity
dispersion is 12 km/s whereas for old LPVs, it reaches 35 km/s.
More recently, Zaritsky \etal (1999) found a velocity dispersion of 
$\sigma = 18.4 \pm 1.4$ km/s for 190 vertical red clump (VRC) 
stars\footnote{see Zaritsky \etal (1999) and Beaulieu and Sackett 
(1998) for a definition of RC and VRC stars.} whereas for the red 
clump (RC), they measured a value of $\sigma = 32.2 \pm 3.8$ km/s on a 
sample of 75 objects (throughout this paper, error bars are converted 
from Zaritsky's 95 \% confidence levels to standard $1 \sigma$).  A 
general trend appears: the velocity dispersion is an increasing 
function of the age.
Just like for our own Milky Way, stars of the \lmc disk have been
continuously undergoing dynamical scattering by, for instance, molecular
clouds or other gravitational inhomogeneities. This results in an increase
of the velocity dispersion of a given stellar population with its
age, as will be further discussed in section 3.
Notice that the main argument
in disfavour of a \lmc self-lensing explanation is precisely the low value
of the measured vertical velocity dispersions.
However, the stellar populations so far surveyed predominantly
consist of red giants. They are shown 
in the next section not to be representative of the bulk of the \lmc disk 
stars, and actually biased towards young ages: 
they are on average $\sim$ 2 Gyr old, to be compared to an \lmc
age of $\sim$ 12 Gyr. 

 %

\section{The age bias}

The red clump population will illustrate the main thrust of our
argument. Clump stars have burning helium cores whose size is
approximately independent of the total mass of the object.
They also have the same luminosity and hence they spend a
fixed amount of time $\tau_{\rm \, He}$ in the clump, irrespective
of their mass $m$. Such objects are evolved post-MS stars,
which does not mean that they are necessarily old.
We have assumed a Salpeter Initial Mass Function for the
various \lmc stellar populations
\begin{equation}
{\displaystyle \frac{dN}{dm}} \propto
m^{\displaystyle - \left( 1 + \alpha \right)} 
\;\; ,
\end{equation}
with $\alpha = 1.35$.
The stellar formation history has been borrowed from
Geha \etal (1998).
Their preferred model (e) corresponds to a stellar formation rate
${\cal F}(t)$ that has remained constant for 10 Gyr since the formation
of the \lmc 12 Gyr ago. Then, two Gyr ago, ${\cal F}(t)$ has increased
by a factor of three. The number of stars that formed at time $t$ and
whose mass is comprised between $m$ and $m + dm$ may
be expressed as
\begin{equation}
{\displaystyle \frac{d^{2} N}{dm \, dt}} = {\cal F}(t) \,
m^{\displaystyle - \left( 1 + \alpha \right)} \;\; .
\end{equation}
We have assumed a mass-luminosity relation $L \propto m^{\beta}$ on 
the MS so that the stellar lifetime may be expressed as $\tau_{\rm MS} 
(m) = {12 \; {\rm Gyr}} / {m^{\beta - 1}}$ (since $\tau \propto m/L$).
With these oversimplified but natural assumptions, a star whose initial
mass is $\leq 1 \, \Msol$ is still today on the MS and cannot have
reached the clump. Conversely, a heavier star with $m \geq 1 \, \Msol$
may well be today in a helium core burning stage provided that its
formation epoch lies in the range between
$t = - \, \tau_{\rm MS} (m)$ (the object has just begun core helium
burning) and
$t = - \, \tau_{\rm MS} (m) - \, \tau_{\rm \, He} (m)$ (the star is about to
leave the red clump). The number of RC stars observed today with
progenitor mass in the range between $m$ and $m + dm$ is therefore
given by
\begin{equation}
dN_{\rm RC}  =  {\cal F} ( - \tau_{\rm MS} (m) ) \times
m^{\displaystyle - \left( 1 + \alpha \right)} \, dm \times \tau_{\rm \, 
He} \;\; .
\label{CHSTAR_1}
\end{equation}

To get more insight into the age bias at stake, we can parameterize the
progenitor mass $m$ in terms of the age
$\tau \equiv \tau_{\rm MS} (m)$. The previous relation simplifies
into
\beq
\frac{dN_{\rm RC}}{d\tau} \; = \;
{\displaystyle
\frac{{\cal F} ( - \tau ) \, \tau_{\rm \, He}}{(\beta - 1)}} \;
\tau^{\displaystyle \left( \gamma - 1 \right)} \;\; ,
\eeq
where $\gamma = \alpha / (\beta - 1)$. This may be directly compared
to the age distribution of the bulk of the \lmc stars that goes like
${\cal F} ( - \tau )$. With a Salpeter mass function
and $\beta = 4.5$, we get a value of $\gamma = 0.4$. The excess
of young RC stars goes as $1 / \tau^{0.6}$ and the bias is obvious.
Other IMF are possible and a spectral index as large as
$\alpha \sim \beta - 1 \sim 3.5$ would be required to invalidate
the effect. HST data analyzed by Holtzman \etal (1997) nevertheless
point towards a spectral index $\alpha$ that extends from 0.6 up to 2.1
for stars in the mass range $0.6 \leq m \leq 3$ $\Msol$. The average
value corresponds actually to a Salpeter law.

There has been furthermore a recent burst in the \lmc stellar formation
rate. In order to model it, we may express the total number of
today's RC stars as an integral where the progenitor mass $m$ runs
from $m_{1} = 1 \, \Msol$ up to the tip of the IMF whose actual value
is irrelevant and has been set equal to infinity here for simplicity.
Notice that the specific progenitor mass
$m_{2} \simeq 1.7 \, \Msol$ corresponds to stars born 2 Gyr ago,
when the stellar formation rate increased by a factor of 3. Stars which
formed before that epoch will be referred to as old. Their number is
given by
\begin{equation}
N_{\rm RC}^{\rm old} =
{\displaystyle \int_{m_{1}}^{m_{2}}} \,
{\cal F} \left( - \tau_{\rm MS} \right) \,
m^{\displaystyle - \left( 1 + \alpha \right)} \, dm
\, \tau_{\rm \, He} \;\; .
\end{equation}
On the other hand, the number $N_{\rm RC}^{\rm young}$ of young
clump stars is obtained similarly, with masses in excess of $m_{2}$.
We readily infer a fraction of young stars
\begin{equation}
{N_{\rm RC}^{\rm young}} / N_{\rm RC} =
{\displaystyle \frac{3}{2 \, + \, (m_{2}/m_{1})^{\alpha}}}
\simeq 0.751 \;\; .
\end{equation}
Three quarters of the clump stars observed today in the \lmc have
thus formed less than 2 Gyr ago, during the recent period of stellar
formation mentioned above. Integrating $\tau_{\rm MS}$
over the RC population
\begin{equation}
\left< \tau \right> = \frac{1}{N_{\rm RC}} \;
{\displaystyle \int_{m_{1}}^{\infty}} \tau_{\rm MS} \,
dN_{\rm RC} \;\; ,
\end{equation}
yields the average age
\begin{equation}
\left< \tau \right>  =  \left( 12 \; {\rm Gyr} \right)  \times 
 \frac{\alpha}{\alpha + \beta - 1}  \times 
 {\displaystyle \frac
{m_{1}^{1 - \alpha - \beta} \, + 2 \, m_{2}^{1 - \alpha - \beta}}
{m_{1}^{- \alpha} \, + 2 \, m_{2}^{- \alpha}}}
 \; .
\end{equation}
This gives a numerical value of $\sim 1.95$ Gyr. We thus
conclude that today's clump stars are, on average, much younger
than the \lmc disk.

\section{Distributions of velocity dispersions}

This simple analytical result has been checked by means of
a Monte Carlo study. We have randomly generated a sample of
$10^{8}$ \lmc stars. The progenitor mass was drawn in the range
$0.1 \leq m \leq 10$ $\Msol$ according to a Salpeter law.
The age of formation was drawn in the range
$-12$~Gyr $\leq t \leq 0$ according to the stellar formation
history ${\cal F}(t)$ favoured by Geha \etal (1998). The vertical
velocity dispersion $\sigma_{W}$ was then evolved in time from
formation up to now according to Wielen's (1977) relation:
\begin{equation}
\sigma_{W}^{2} \; = \; \sigma_{0}^{2} + C_{W} \, t.
\label{diffeq}
\end{equation}
This purely diffusive relation is known to be inadequate to describe 
velocity dispersions in our Galaxy (Edvardsson et al. 1993).  We will 
however use it in our model, as heating processes in the \lmc may be 
different than those in the galaxy.  The \lmc is indeed subject to 
tidal heating by the Milky Way (\cite{weinberg99}) and has most 
probably suffered encounters with the \smc .  Although this simple 
relation lacks a theoretical motivation, it will be shown to account 
for several features of the velocity distributions in the \lmc, 
without being at variance with any observation.
The initial velocity dispersion $\sigma_{0}$ was taken to be 10~km/s,
and the diffusion coefficient in velocity space along the vertical
direction $C_{W}$ to be
300~$\unit{km}^{2} \unit{s}^{-2} \unit{Gy}^{-1}$
so that our oldest stars have a vertical velocity dispersion reaching
up to $\sigma_{W}^{\rm MAX} = 60$ km/s. For each star, the actual
vertical velocity was then randomly drawn, assuming a Gaussian
distribution with width $\sigma_{W}$.

In order to compare our Monte Carlo results with the Zaritsky \etal
(1999) measurements of the radial velocities of \lmc clump stars, we
selected two groups of stars according to their position in the HR
diagram. Following Zaritsky \etal, we use their colour index
\beq
C \; \equiv \; 0.565 \, (B - I) \; + \; 0.825 \, (U - V + 1.15) \;\; ,
\eeq
so that the RC population is defined by $3.1 < C < 3.4$ with a magnitude
$19 < V < 19.3$ whereas the VRC stars have the same colour index C
and brighter magnitudes $18 < V < 18.75$. In order to infer the
colours and magnitudes of the stars that we generated, we used the
isochrones computed by Bertelli \etal (1994) for a typical \lmc
metallicity and helium abundance of $Z = 0.008$ and $Y = 0.25$.

\begin{figure}[h!]
\begin{center}
\epsfig{file=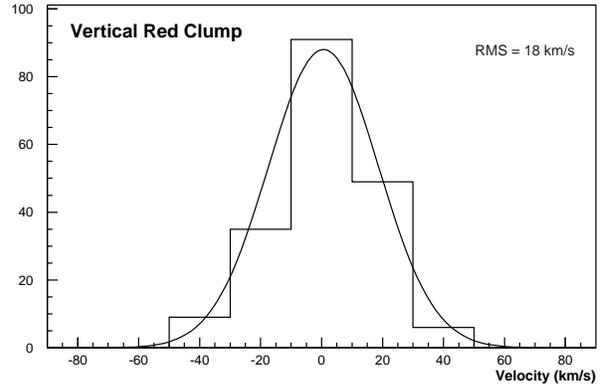,width=7.8cm}
\caption{
Velocity distribution for a sample of 190 vertical
red clump stars that have been generated by the Monte Carlo
discussed in the text. That histogram is similar to Fig.~10 of
Zaritsky \etal (1999). A velocity dispersion of 18~km/s is found
for the full sample (solid smooth curve).
}
\label{fig_vrc}
\end{center}
\end{figure}

A random sample of 190 stars that passed the VRC selection criteria is presented
in Fig.~\ref{fig_vrc} where the vertical velocities are displayed. This
histogram may be compared to Fig.~10 of Zaritsky \etal (1999) where
no VRC star is found with a velocity in excess of 60 km/s. With the full
statistics, our Monte Carlo generated a population of $\sim$ 2,900 VRC
objects whose vertical velocity distribution has a RMS of $\sim$ 18
km/s. The agreement between the Zaritsky \etal observations and our
Monte Carlo results is noteworthy. The average age of our VRC sample
is $\sim$~0.87~Gyr.

\begin{figure}[h!]
\begin{center}
\epsfig{file=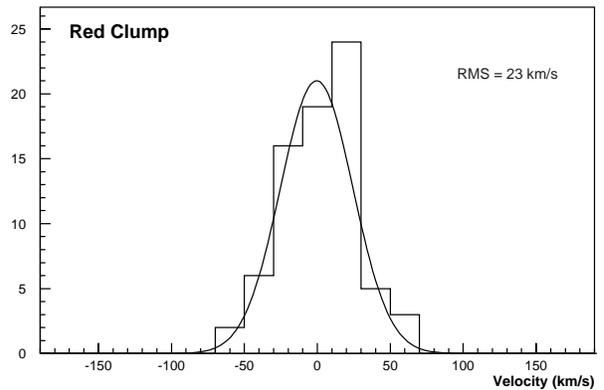,width=7.8cm}
\caption{
Like in the previous figure, a distribution of 75 red clump stars
is now featured. We inferred a velocity dispersion of 23 km/s
for the full sample (solid smooth curve). Our distribution is similar to that presented
in Fig.~11 of Zaritsky \etal (1999).
No star exhibits a velocity larger than 70~km/s.
}
\label{fig_rc}
\end{center}
\end{figure}

We also selected a random sample of 75 RC stars whose velocity distribution
is featured in Fig.~\ref{fig_rc}. Even with a diffusion coefficient as
large as $C_{W} = 300 \unit{km}^{2} \unit{s}^{-2} \unit{Gy}^{-1}$
so as to comply with a large \lmc self-lensing optical depth, our full
statistics of 18,000 RC objects has a velocity dispersion of $\sim$ 23 km/s.
This is slightly below the value of $\sigma = 32.2 \pm 3.8$ km/s quoted
by Zaritsky \etal
Observations are nevertheless fairly scarce with only 75 RC stars.
When Zaristsky \etal fitted a Gaussian to the RC radial velocity distribution
featured in the Fig.~11 of their paper, they obtained a 95~\% C.L. dispersion of
$\sigma = 32^{+19}_{-16}$ km/s with a large uncertainty. Our Monte Carlo
velocity dispersion of 23 km/s is definitely compatible with that result.
We infer
an average age for the RC population of $\sim$~1.8 Gyr to be compared
to our analytical result of $\sim$ 1.95 Gyr. This agrees well with Beaulieu
and Sackett's conclusion that isochrones younger than 2.5 Gyr are
necessary to fit the red clump. Notice finally that our age estimates
for these various clump populations are in no way related to \lmc
kinematics. They merely result from the postulated Salpeter IMF,
the Geha \etal preferred stellar formation history and the
Bertelli \etal isochrones.

With this model, 70\% in mass of the \lmc disk consists of objects 
whose vertical velocity dispersion is in excess of 25 km/s, although 
the average vertical velocity dispersion of RC stars, for instance, is 
only $\sim$ 23 km/s.

What about the other measurements?
%
%
The velocity dispersion of PNs has been found equal to 19.1 km/s
(Meatheringham \etal 1988). These authors estimate that the bulk
of the PNs have an age near 3.5 Gyr. They also note that younger
objects are present down to an age of order $0.5 - 1.3$ Gyr.
Meatheringham \etal come finally to the conclusion that the
indicative age of the PN population is 2.1 Gyr. This value agrees well
once again with our analytical estimate. Our Monte Carlo gives a
slightly larger value of 2.4 Gyr for the age of the PNs, with a velocity
dispersion of 24.7 km/s. Because the observed sample contains 94 objects,
the measured value of 19.1 km/s suffers presumably from significant
uncertainties.
%
%

Quite interesting also are the measurements by Hughes \etal (1991) of 
the velocity dispersions of LPVs as a function of their age.  Their 
sample of 63 ``old'' LPVs has a velocity dispersion of $\sigma = 
35^{+10}_{-4}$ km/s.  For the bulk of the \lmc populations, we obtain 
an average velocity dispersion of $\sim 37$ km/s.  The problem at 
stake is actually the age of those old LPVs.  These stars indeed 
display an age-period relation.  However, Hughes \etal derived this 
relation from kinematics considerations, using precisely Eq.  
\ref{diffeq}, and postulating the same diffusion coefficient as in the 
Milky Way.  They thus inferred an average age of 9.5~Gyr.  Finding 
instead the position of these stars in a colour-magnitude diagram and 
using \lmc isochrones would have led to a clean determination of the 
age-period relation.
A direct determination of the age of LPVs is nevertheless spoilt by a 
few biases.  Some LPVs are carbon stars and the ejected material 
around them may considerably dim their luminosities.  These stars may 
also pulsate on an harmonic of the fundamental mode.  Both effects 
lead to an under-determination of their luminosity and hence to an 
overestimate of their age (Menessier 1999).  As a matter of fact, 
Groenewegen and de Jong (1994) conclude that \lmc stars whose 
progenitor mass is less than 1.15 $\Msol$ never reach the instability 
strip on the AGB. This yields an upper limit on the age of LPVs of 
$\sim 7.3$ Gyr, in clear contradiction with the average age of 9.5 Gyr 
inferred by Hughes \etal for old LPVs.
%
%

Finally, Schommer \etal (1992) have obtained a velocity dispersion
of $21 - 24$ km/s for 9 old \lmc clusters. Their large $1\sigma$ error
of $\sim$ 10 km/s is due to the small size of the sample.
It is not clear whether or not these clusters have formed in the disk.
If they nevertheless had, they would have undergone a fairly
restricted orbital heating with respect to the \lmc stars. Those
systems and the giant molecular clouds have actually comparable
masses and the energy exchange between them does not result
in a significant increase of the velocity dispersion of the clusters
unlike what happens to the stars.


\section{Multi-component model of the \lmc}

We model the \lmc to contain several stellar populations, each
associated with a different velocity dispersion $\sigma_{W,i}$
which has evolved according to Eq. \ref{diffeq}.

We describe each of the ten components of our model by an ellipsoidal
density profile
\begin{equation}
\rho_i(R,z) =
\frac{\Lambda_i}{R^2 \; + \; {\displaystyle {z^2}/{(1-e_i^2)}}}
\;\; ,
\end{equation}
up to a cut-off radius $R_{\rm MAX} = 15 \unit{kpc}$ (Aubourg et al.
1999).  The multi-component model based on these profiles is
self-consistent in the sense that it satisfies Poisson equation and
results in a flat rotation curve with the desired $V_C$ of 80~km/s.
We define the set of $\sigma_{W,i}$ so as to sample linearly the range
between $\sigma_0 = 10 \unit{km/s}$ and $\sigma_{W}^{\rm MAX} = 60
\unit{km/s}$ (see previous section).  The parameters $\Lambda_i$ and the ellipticities $e_i$
are determined so that the model reproduces the set of velocity
dispersions $\sigma_{W,i}$ and surface mass densities $\Sigma_i$ where
$d\Sigma_i/d\sigma_i \propto \sigma_i {\cal F}(t)$ with ${\cal F}(t)$
the stellar formation history of the \lmc mentioned in section 2.
Assuming a typical M/L of 3, which is a free parameter in our model,
we reproduce the observed surface brightness of the LMC.

For a given distribution of objects, one can compute the total
self-lensing optical depth $\tau$ and the event rate $\Gamma$.
Both quantities are integrated on all deflectors and sources, considering
that only main sequence stars brighter than $V = 20$ and red giants can
be potential sources, since they are the only objects bright enough to be
visible in microlensing surveys. The computation of $\Gamma$ requires
an estimate of the relative transverse velocity of deflector and source, for
which we have assumed an horizontal velocity dispersion equal to the
vertical one predicted by the model. Details of this computation can be
found in (\cite{nous}).

For the model described above, one obtains $\tau = 9.3 \times 10^{-8}$
and $\Gamma = 3.5 \times 10^{-7} \unit{yr}^{-1}$.
This can be compared to the \eros and \macho optical depths,
respectively
$8.2 \times 10^{-8}$ (\cite{eros1}) and
$2.9_{-0.9}^{+1.4} \times 10^{-7}$ (Alcock \etal 1997). A combination
of those two results yields an average optical depth of
$2.1_{-0.8}^{+1.3} \times 10^{-7}$ (\cite{BennettReport}), but preliminary
\macho results from their five-year analysis (Sutherland 1999) hint to a
reduced optical depth as compared to their two-year analysis. The model
prediction is thus in good agreement with the results obtained so far from
microlensing experiments.

\begin{figure}[h]
\begin{center}
\epsfig{file=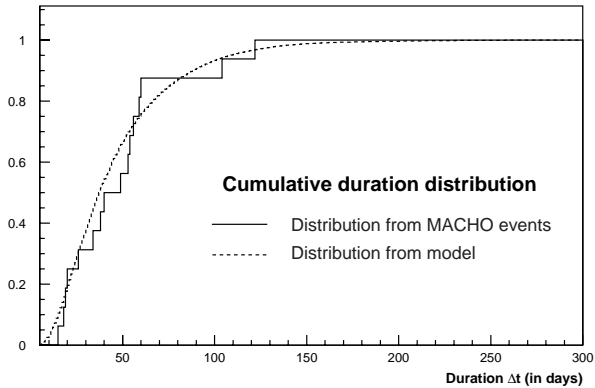,width=7.8cm}
\caption{
Predicted distribution of event durations $d\Gamma/d\Delta t$,
superimposed with the \macho experimental distribution. The
events are those presented by \macho at the IVth Microlensing
Workshop (\cite{machoEvt}), corrected for blending and efficiency
using the formulae in \cite{Macho2yr}.
}
\label{dgamdtau}
\end{center}
\end{figure}

Another relevant prediction of the model is the distribution of event
durations, $d\Gamma/d\Delta t$. Figure~\ref{dgamdtau} illustrates this
prediction for our model, along with the distribution of observed
\macho events.

Our model thus reproduces both the total observed optical depth towards
the \lmc and the observed event duration distribution, while complying
with the velocity dispersion measurements. A self-lensing interpretation
of {\em all} the microlensing events observed so far towards the \lmc
thus appears to be a plausible explanation.

\begin{acknowledgements}
We wish to thank M.O. Menessier for useful 
discussions, and the members of the EROS collaboration for their comments.
We thank Andy Gould, our referee, for his useful remarks and suggestions.
\end{acknowledgements}

\end{document}